\global\newcount\secno \global\secno=0
\def\newsec#1{\global\advance\secno by1%
\global\meqno=1%
\bigskip\noindent{\bf\the\secno. #1}\par\nobreak\medskip\nobreak}

\global\newcount\meqno \global\meqno=1
\def\eqn#1#2{\xdef#1{(\the\secno.\the\meqno)}%
$$#2\eqno#1$$%
\global\advance\meqno by1}

\global\newcount\refno \global\refno=1
\newwrite\rfile
\immediate\openout\rfile=refs.tmp
\def\ref#1#2{\xdef#1{\the\refno}%
\immediate\write\rfile%
{\medskip\noexpand\item{[#1]}\noexpand{#2}}%
\global\advance\refno by1}

\def\foot#1#2{\footnote{$\!^{#1}$}{#2}}

\def\ack#1{\bigskip\noindent{\bf Acknowledgments}
\par\nobreak\medskip\nobreak#1}


\pageno=0
\footline={\ifnum\pageno>0 \hfil\number\pageno\hfil \else\hfil \fi}

\def\abstract#1{{\noindent\narrower#1\par}}

\def\arcsinh{\hbox{arcsinh $\!$}}

\def\inprod#1#2{\bigl<#1\,\,\big\vert\,\,#2\bigr>}
\def\pd#1#2{{\partial #1 \over \partial #2}}

\def\prug{\noexpand{Prugove\v{c}ki}}
\def\rhonull{{\rho_{\rm null}}}
\def\rhosl{{\rho_{\rm sl}}}
\def\rmax{{r_{\rm max}}}
\def\Ctl{C_{\rm tl}}
\def\Csl{C_{\rm sl}}

\def\BC{{\bf C}}
\def\BE{{\bf E}}
\def\BF{{\bf F}}
\def\BK{{\bf K}}
\def\BM{{\bf M}}

\def\BR{{\bf R}}
\def\BV{{\bf V}}
\def\BX{{\bf X}}

\def\B0{{\bf 0}}

\def\Ba{{\bf a}}

\def\Be{{\bf e}}
\def\Bg{{\bf g}}
\def\Bk{{\bf k}}
\def\Bn{{\bf n}}
\def\Bp{{\bf p}}
\def\Bq{{\bf q}}
\def\Bs{{\bf s}}
\def\Bu{{\bf u}}
\def\Bx{{\bf x}}

\magnification=\magstep1

\ref\rprugA {\prug\ E 1995 {\it Principles of Quantum General 
  Relativity} (Singapore : World Scientific)}
\ref\rprugE {\prug\ E 1996 {\it Quantum Gravity} eds P G Bergmann,
  V de Sabbata, and H J Treder (Singapore : World Scientific) pp 276-302}
\ref\rbyt {Bytsenko et al 1996 {\it Phys. Rep.} {\bf 266} 1}
\ref\rbjork {Bjorken J D and Drell S D 1965 {\it Relativistic
  Quantum Mechanics} (New York : McGraw-Hill)}
\ref\rheg {Hegerfeldt G C 1974 {\it Phys. Rev. D} {\bf 10} 3320}
\ref\rwig {Wigner E P 1983 {\it Quantum Theory of Measurement} eds
  J A Wheeler and W M Zurek (Princeton : Princeton University Press) 
  pp 310-314}
\ref\rheis {Heisenberg W 1930 {\it Z. Phys.} {\bf 43} 4}
\ref\rbornA {Born M 1938 {\it Proc. Roy. Soc. London A} {\bf 165} 291} 
\ref\rdewitt {DeWitt B S 1962 {\it Gravitation : An Introduction to Current
  Research} ed L Witten (New York : Wiley)} 
\ref\rvon {von Borzeszkowski H H and Treder H J 1988 {\it The Meaning of 
  Quantum Gravity} (Dordrecht : Reidel)} 
\ref\rprugD {\prug\ E 1984 {\it Stochastic Quantum Mechanics and 
  Quantum Spacetime} (Dordrecht : Reidel)}
\ref\rschroeck {Schroeck F E Jr 1995 {\it Quantum Mechanics on Phase
  Space} (Dordrecht : Kluwer)}
\ref\rbornB {Born M 1949 {\it Rev. Mod. Phys.} {\bf 21} 463} 
\ref\rbornC {Born M 1939 {\it Proc. Roy. Soc. Edinburgh} {\bf 59} 219}
\ref\rlande {Land\'e A 1939 {\it Phys. Rev.} {\bf 56} 482,486}
\ref\rali {Ali S T and \prug\ E {\it Nuovo Cimento A} {\bf 63} 171}
\ref\rdrecA {Drechsler W and Tuckey P 1996 {\it Class. Q. Grav.} 
  {\bf 13} 611}
\ref\rprugB {\prug\ E 1996 {\it Class. Q. Grav.} {\bf 13} 1007}
\ref\rkob {Kobayashi S and Nomizu K 1963 {\it Foundations of Differential
  Geometry} vol 1 (New York : Wiley)}  
\ref\rdrecB {Drechsler W 1984 {\it Fortschr. Phys.} {\bf 23} 449}
\ref\rspivak {Spivak M 1979 {\it A Comprehensive Introduction to Differential
  Geometry} vol 2 (Wilmington : Publish or Perish)}
\ref\rschutz {Schutz B F 1985 {\it A First Course in General Relativity} 
  (Cambridge : Cambridge University Press)}
\ref\rprugC {\prug\ E private communication}
\ref\rdods {Dodson C T J and Poston T 1991 {\it Tensor Geometry} 
  (Berlin : Springer)}

\centerline {\bf ON PARALLEL TRANSPORT IN QUANTUM BUNDLES}
\centerline {\bf OVER ROBERTSON-WALKER SPACETIMES}
\bigskip
\bigskip
\bigskip
\bigskip
\centerline {{\bf James Coleman} \foot\dag {coleman@math.utoronto.ca}}
\bigskip
\centerline {Department of Mathematics}
\centerline {University of Toronto}
\centerline {Toronto, Canada M5S 3G3}
\bigskip
\bigskip
\bigskip
\bigskip
\bigskip
\bigskip

\abstract {{\bf Abstract.} A recently-developed theory of quantum general 
relativity provides a propagator for free-falling  particles in curved 
spacetimes.  These propagators are constructed by parallel-transporting 
quantum states within a quantum bundle associated to the Poincar\'e frame 
bundle.  We consider such parallel transport in the case that the spacetime 
is a classical Robertson-Walker universe.  An explicit integral formula is
developed which expresses the propagators for parallel transport
between any two points of such a spacetime.  The integrals in this formula 
are evaluated in closed form for a particular spatially-flat model.}

\bigskip

\abstract {PACS numbers : 0420, 0240, 0220}

\vfill\eject

\newsec {Introduction}

In a recent book [\rprugA], \prug\ describes in detail a framework for the
unification of general relativity and geometro-stochastic quantum 
theory (cf.\ [\rprugE] for a review).
The theoretical basis for geometro-stochastic quantization
differs from the orthodox quantization procedure in that it
incorporates the concept of a fundamental length in nature,
which limits the accuracy with which one can measure the spacetime
location of an event.  This assumption resolves many of the difficulties
with divergences encountered in conventional quantum field theory.

In this paper, we focus on the propagation of quantum wave functions for a 
massive spinless boson in a Lorentzian manifold $\BM$.  In particular, we 
derive the equations governing such propagation in the case that this 
manifold is a classical Robertson-Walker model of our universe.  The 
methods described could easily be applied to other models that have
received attention in [\rbyt] in the context of conventional field
theory in curved spacetimes. 

Throughout this paper,  we assume that $\BM$ is space-time orientable, so
that it can be covered by a global coordinate system of orthonormal
frames in which the timelike vector is future-pointing and the spacelike
vectors form a right-handed triad.  In addition, there is assumed to be
a global time variable $t$ so that we can foliate $\BM$ into a set of
Cauchy spacelike hypersurfaces $\sigma_t$.
We adopt a metric $\Bg$ with the signature $(-1,1,1,1)$ commonly used 
in relativity theory, and use Planck natural units where $c = G = \hbar = 1$.
 
In section 2 we provide a brief overview of the special-relativistic case 
in which a single wave function on phase space is used to 
describe the state of the boson.  This is then extended in section 3
to the case of a globally hyperbolic spacetime through the introduction of 
the general quantum bundle, which is associated to the Poincar\'e frame 
bundle over that spacetime.  Parallel transport on this bundle is defined by
making use of the standard extension of the Levi-Civita connection used in 
classical general relativity, and serves as the basis for the development of
the quantum-geometric propagator.  In section 4, we compute the geodesics 
for the Robertson-Walker model and use them to develop an integral 
representation of the propagators for parallel transport.  Finally, in 
section 5 we focus on a particular spatially-flat model for which these 
propagators turn out to have a closed-form expression.

\newsec {Massive boson states in special relativistic phase space}

Consider a spinless particle of rest mass $m > 0$ which moves freely in 
the Minkowski space $M^4$.  The momentum representation of such a particle 
consists of the space $L^2(\BV_m^+)$ of the standard wave functions on
the forward mass hyperboloid corresponding to $m$, whose inner
product is expressed using the usual invariant measure on $\BV_m^+$: 
\eqn\eVmplusdef{
{\eqalign{&\inprod {\tilde\varphi_1}{\tilde\varphi_2}=
\int_{\BV_m^+}\tilde\varphi_1^*(k)\tilde\varphi_2(k)\,d\Omega_m(k),
\quad \tilde\varphi_1,\tilde\varphi_2 \in L^2(\BV_m^+), \cr
\BV_m^+ = \bigl{\{}k &\in M^4,\,|\,k^0 = \sqrt{\Bk^2+m^2}\bigr{\}}, \quad 
d\Omega_m(k) = {d\Bk \over 2 \sqrt{\Bk^2 + m^2}},\quad
k = (k^0,\Bk).}}}
As is well-known [\rbjork], the standard interpretation of $\tilde\varphi$
is that of a probability amplitude for particle 4-momentum measurements.
However, a corresponding relativistically invariant probability amplitude
does not exist for spacetime measurements [\rheg,\rwig].

Already in the 1930's, various physicists, of whom the most prominent were
Heisenberg [\rheis] and Born [\rbornA], had conjectured that
there is a fundamental length $\ell$ in nature, such
that no spacetime measurement of an event can be made with greater
accuracy than $\ell$.  Measurement-theoretical arguments [\rdewitt,
\rvon] suggest that this length is in fact Planck's length (which is 1 
in the natural system of units adopted in this paper).

In the quantum-geometric framework,
this hypothesis forms an integral part of the structure of the 
single-boson Hilbert space $\BF$ which represents the state of the 
system.  In the remainder of this section we present an overview of this 
theory; for details consult sections 3.7 and 3.8 of [\rprugA] and chapter 
2 of [\rprugD].

To construct $\BF$, we first introduce the concept of
the momentum space resolution generator, denoted by $\tilde\eta$.
{}From a mathematical viewpoint, this resolution generator can be
chosen to be any wave function from $L^2(\BV_m^+)$ satisfying the 
following two properties; first that $\inprod {\tilde\eta} {\tilde\eta} =
{m / 4 \pi^3}$, and second that $\tilde\eta$ is invariant under spatial 
rotations.
 
To understand the physical significance of $\tilde\eta$, it is necessary
to consider in detail the actual process of measuring the spacetime
coordinates of a given event relative to some inertial frame.
Classically, we can imagine that this process is performed with
the aid of an apparatus employing pointlike particles
and light signals.  However, a closer examination of this procedure shows 
that the recoil effects produced by the
emission and absorption of photons will introduce an inherent uncertainty 
factor.  If, however, we replace the pointlike particles 
by extended particles whose momentum space wave function is given
by $\tilde\eta$, then it is possible to define this process in a physically 
consistent manner which takes advantage of the various methods developed 
in the phase space approach to quantum mechanics [\rschroeck].
   
The introduction of quantum frames of reference induces a fundamental
change in the notion of spacetime itself.  Specifically, the position and
momentum coordinates $\Bq \in \BR^3$ and $\Bp \in \BR^3$ of the particle
which are measured at some given time will now be {\it stochastic} variables 
(cf.\ section 3.3 of [\rprugA]) whose definitions incorporate Gaussian-type 
distributions $\chi_\Bq$ and $\chi_\Bp$ on $\BR^3$.   The standard deviations 
of these distributions are of the order of $\ell$ and $\ell^{-1}$ respectively
(in Planck units).  Hence, it is possible to define
a wave function on the phase space of the particle which provides
a joint probability amplitude for measuring the particle as
simultaneously having a given stochastic position and stochastic 
momentum, without violating the Heisenberg uncertainty principle.
This will be done below.  First, however, we consider what value
the momentum space resolution generator should actually have. 

In general, there are infinitely many possible choices for $\tilde\eta$. 
However, in this paper we will henceforth consider only the 
special case
\eqn\emomresval{
\tilde\eta(k) = \bigl(m^3 \tilde Z_{\ell,m}\bigr)^{-{1 \over 2}}
\exp(-\ell k^0), \quad 
\tilde Z_{\ell,m} = {8 \pi^4 \over \ell m^2}\,K_2(2 \ell m),
\quad k=(k^0,\Bk) \in \BV_m^+,}
where $K_2$ denotes as usual a modified Bessel function.  This case 
corresponds to the unique resolution generator whose phase space 
representative displays reciprocal invariance under transposition of 
configuration and momentum coordinates, and is the ground state 
eigenvector of Born's [\rbornB] quantum metric operator. 
Hence, its presence reflects the fact that stochastic spacetime 
localization is implemented only to the order of $\ell$, whereas the
corresponding stochastic 4-momentum spread is of the optimal order of
magnitude that is in accordance with the Heisenberg uncertainty
principle [\rschroeck].

The above choice of $\tilde\eta$ induces a mapping 
${\cal W}_\eta$ from $L^2(\BV_m^+)$ into the set of complex-valued 
functions defined on the relativistic phase 
space ${\cal P}_m^+ = \{(q,p) \in M^4 \times \BV_m^+\}$, given by
\eqn\eWdef{
{\cal W}_\eta : \tilde\varphi \mapsto \varphi, \quad \varphi(q,p) =
\inprod {\tilde\eta_{q,p}} {\tilde\varphi}, \quad \tilde\eta_{q,p}(k) = 
\exp(-i q \cdot k)\,\tilde\eta(\Lambda_v^{-1}k),}
where $\Lambda_v$ is the Lorentz boost corresponding to the 4-velocity
$v = p/m$.  The image under ${\cal W}_\eta$ of $L^2(\BV_m^+)$ forms a 
Hilbert space $\BF$ with inner product
\eqn\einprodF{
\inprod {\varphi_1}{\varphi_2}=
\int_{\Sigma_m^+}\varphi_1^*(q,p)\varphi_2(q,p)\,d\Sigma_m(q,p),
\quad\quad \varphi_1,\varphi_1 \in \BF.}
The domain of integration is the 6-dimensional hypersurface $\Sigma_m^+ = 
\sigma \times \BV_m^+$, where $\sigma$ is any Cauchy spacelike hypersurface 
in $M^4$.  The hypersurface $\Sigma_m^+$ carries the relativistically 
covariant volume element 
\eqn\edSigmadef{
d\Sigma_m(q,p) = 2p^i\,d\sigma_i(q)\,d\Omega_m(p),\quad\quad
(q,p) \in \Sigma_m^+.}
It can be shown [\rprugD,\rschroeck] that the conditions imposed on the 
resolution generator $\tilde\eta$ imply that the mapping ${\cal W}_\eta$ 
defined in \eWdef\ is unitary.  Consequently, for normalized $\varphi$
the values assumed by the probability measure
\eqn\eprob{
P_\varphi(B) = \int_B \vert\varphi(q,p)\vert^2\,d\Sigma_m(q,p)}
can be interpreted as representing relativistically invariant
probabilities for stochastic phase space localization.

This phase space representation has several advantages over the
conventional formulation, which is usually based on wave functions
defined on $M^4$ that satisfy the Klein-Gordon equation.
In addition to providing a joint probability amplitude for
stochastic position and momentum as explained above, $\varphi$ 
simultaneously satisfies the Klein-Gordon equation in its $q$-variables
and the Born-Land\'e equation [\rbornC,\rlande] in its $p$-variables:
\eqn\eKGBL{
\left(\eta^{ij}{\partial^2 \over q^i q^j}-m^2\right)\varphi(q,p) = 0, 
\quad\quad \left(\eta^{ij}{\partial^2 \over p^i p^j}-\ell^2
\right)\varphi(q,p) = 0.}
Moreover, on the physical side, $\varphi$ can be used to define a
positive-definite, conserved probability current, so that a
physically-consistent single-particle theory becomes possible.
This theory can also be extended to handle particles in external
electromagnetic fields.
For example, it is possible to develop a model in which the
particle is coupled to a stochastic version of the usual 4-potential $A_i$, 
for both the Klein-Gordon equation (section 2.10 of [\rprugD]), and the 
Dirac equation [\rali].  The time development of the wave function in such 
models does not give rise to transitions to negative energy states.  Hence, 
as shown in detail in [\rprugD] and [\rschroeck], inconsistencies 
such as the Klein paradox are avoided.

The phase space ${\cal P}_m^+$ is acted upon by the restricted Poincar\'e
group $ISO_0(3,1)$.  Each element of $ISO_0(3,1)$ can be represented by a
pair $(b,\Lambda)$, where $b$ is a vector in $\BR^4$ and $\Lambda$
is a 4-by-4 matrix representing a Lorentz transformation. 
The action of $(b,\Lambda)$ on points in ${\cal P}_m^+$ is given by
\eqn\ePoinactionP{
(b,\Lambda)(q,p) = (\Lambda q + b, \Lambda p), 
\quad\quad (b,\Lambda) \in ISO_0(3,1),}
and has the standard composition law $(b,\Lambda)(b',\Lambda') = 
(b + \Lambda b', \Lambda\Lambda')$.  In turn, this action induces a 
representation $U$ of $ISO_0(3,1)$ on $\BF$.  This representation is
defined  by 
\eqn\ePoinrepH{
U(b,\Lambda) : \varphi(q,p) \mapsto  \varphi\bigl(\Lambda^{-1}(q-b)\,,\,
\Lambda^{-1}p\bigr), \quad\quad (b,\Lambda) \in ISO_0(3,1),}
and physically can be interpreted as providing the relationship between two 
coordinate wave functions corresponding to a particular quantum state, as 
measured by an arbitrary pair of inertial observers in $M^4$ whose relative 
motion is determined by $b$ and $\Lambda$.
 
The resolution generator defined in \emomresval\ can be used to construct a 
special relativistic phase space propagator $K$, which is analogous to the 
Feynman spacetime propagators conventionally used in quantum mechanics.  
To do this, we first introduce the phase space resolution generator 
$\eta \in \BF$, which is defined in terms of \emomresval\ and 
\eWdef\ by $\eta = {\cal W}_\eta\tilde\eta$.  From the rotational
invariance of $\tilde\eta$ it follows that $\eta$ itself satisfies a similar
property, i.e. that $U(0,\Lambda_R)\eta = \eta$ for any spatial rotation
$\Lambda_R \in SO_0(3,1)$.

We now define the quantum propagator as follows.  Let $(q',p')$ and 
$(q'',p'')$ be any two points in ${\cal P}_m^+$.  Then the propagator 
between these points is given in terms of the inner product \einprodF\ 
on $\BF$ by
\eqn\epropdef{
K(q'',p'';q',p') = \inprod {\eta_{q'',p''}} {\eta_{q',p'}}, \quad
\eta_{q'',p''} = U(q'',\Lambda_{v''})\eta, \quad
\eta_{q',p'} = U(q',\Lambda_{v'})\eta.} 
Using \epropdef\ it can then be shown that $K$ satisfies all the 
properties conventionally required of propagators.  In particular, if 
$(q',p')$ and $(q'',p'')$ are arbitrary points in ${\cal P}_m^+$, then 
we have 
\eqn\eKreverse{
K(q'',p'';q',p') = K^*(q',p';q'',p''),}
and
\eqn\eKcomp{
K(q'',p'';q',p') = \int_{\Sigma_m^+}K(q'',p'';q''',p''')\,
K(q''',p''';q',p')\,d\Sigma_m(q''',p'''),}
where, as before, $\Sigma_m^+$ is any allowable hypersurface in ${\cal P}
_m^+$.  From \eKcomp\ it follows that $K$ is a reproducing kernel for 
$\BF$, so that the probability amplitudes corresponding to arbitrary 
phase points in ${\cal P}_m^+$ are related by
\eqn\eKrep{
\varphi(q'',p'') = \int_{\Sigma_m^+}K(q'',p'';q',p')\,\varphi(q',p')
\,d\Sigma_m(q',p'), \quad\quad \varphi \in \BF.}

Although the quantum propagator enjoys all the usual properties of
conventional propagators, it has the advantage of being an analytic
function as opposed to a distribution.  Indeed, for the 
specific choice of momentum resolution generator given in \emomresval, it 
turns out that the propagator \epropdef\ can be evaluated in the closed form
\eqn\epropval{
K(q'',p'';q',p') \,=\,
{2\pi \over m^2 \tilde Z_{\ell,m}}\,{K_1\bigl(m\,f_{\ell,m}(q,p)\bigr) \over 
f_{\ell,m}(q,p)},
\quad q=q''-q', \quad p=p''+p',} 
where $K_1$ again denotes a modified Bessel function, and $f_{\ell,m}$ is 
given by the Poincar\'e invariant expression
\eqn\efmldef{
f_{\ell,m}(q,p) = \sqrt{
q \cdot q - {2i\ell \over m}\,q \cdot p -{\ell^2 \over m^2}\,p \cdot p},
\quad\quad q,p \in M^4.}
This result enables us to manipulate propagators algebraically, without
having to deal with any of the difficulties which arise from attempting
to multiply distributions.  In particular, it will be shown in the
next section that the special relativistic propagator can be used to 
construct a quantum-geometric propagator for curved spacetimes.

\newsec {Quantum propagators in globally hyperbolic spacetimes}

In section 2 we constructed a model of quantum mechanics in special 
relativistic phase space.
The system in consideration was represented by a single wave function 
$\varphi$ defined on the phase space ${\cal P}_m^+$.  In a general curved
spacetime $\BM$ this is no longer possible in a relativistically 
consistent manner.  Instead, it is necessary to
construct a representation based on various principal and associated bundles 
over $\BM$ (cf.\ sections 2.3-2.6 of [\rprugA]).

Within this context, the set of all possible local states of single-boson 
quantum systems will be represented by a bundle $\BE$ associated to the 
Poincar\'e frame bundle $P\BM$, with its typical fibre the Hilbert space 
$\BF$ defined in section 2, and with gauge group $ISO_0(3,1)$ acting 
on $\BF$ in accordance with \ePoinrepH.  

This associated bundle is most easily constructed through the use of
a $G$-product, in terms of which it can be written as $\BE = P\BM \times_G 
\BF.$
This construction implicitly defines a projection map $\pi : \BE \to \BM$
which maps each local quantum state in $\BE$ into its corresponding
base point in $\BM$.  For any $x \in \BM$, the fibre $\pi^{-1}(x)$ above
$x$ is associated to the typical fibre $\BF$ by the soldering map
[\rprugA,\rdrecA,\rprugB]
\eqn\edefsolderH{
\sigma_x^\Bu : \pi^{-1}(x) \to \BF, \quad\quad\quad
{\bf \Psi}_x \mapsto {\it\Psi}_x^\Bu, 
\quad\quad \Bu=(\Ba,\Be_i) \in \Pi^{-1}(x),}
where $\Pi$ is the canonical projection map from $P\BM$ to $\BM$ which 
identifies the base point in $\BM$ of any Poincar\'e frame from $P\BM$.
As usual, the local Poincar\'e frame $\Bu$ has been decomposed into a 
displacement vector $\Ba \in T_x\BM$ and a local Lorentz frame 
$(\Be_0,\Be_1,\Be_2,\Be_3) \in (T_x\BM)^4$.

The physical interpretation of these maps is as follows.  The state of a 
particular system corresponds to a section ${\bf \Psi}$ in $\BE$,
whose value is given by the quantum-geometric propagator described below.
Above each point $x$ in $\BM$ is a local quantum state vector ${\bf \Psi}_x$
which represents the state of the system at $x$.  The coordinate wave 
function $\Psi_x^\Bu : {\cal P}_m^+ \to \BC$ of this state as measured by an 
observer at $x$ is determined by applying to ${\bf \Psi}_x$ the soldering 
map defined in \edefsolderH\ with $\Bu = (\B0,\Be_i)$, where $(\Be_i)$ is 
the local Lorentz frame relative to which the measurements are performed.

Since the quantum bundle $\BE$ is associated to $P\BM$, we can
use the standard extension [\rkob,\rdrecB] to $P\BM$ of the Levi-Civita 
connection on $L\BM$ to construct a connection on $\BE$.  The connection on 
$P\BM$ can be specified by means of the usual 1-forms for infinitesimal
Poincar\'e transformations, whose values in terms of the chosen Poincar\'e 
gauge provided by a section $\Bs$ of $P\BM$ are given by [\rprugB]
\eqn\eformsval{
{\eqalign{
\omega_{ij}^\Bs(\BX) = &{1 \over 2} X^k \bigl[
\Bg(\Be_i,[\Be_j, \Be_k])+
\Bg(\Be_j,[\Be_k, \Be_i])-
\Bg(\Be_k,[\Be_i, \Be_j])\bigr],\cr
&\theta_\Bs^i(\BX) = X^i+\Be^i\bigl(\nabla_\BX\Ba\bigr),
\quad\quad \BX = X^i\Be_i \in T\BM.}}}
By standard results of differential geometry [\rkob], this
connection induces a corresponding connection on $\BE$, which we
refer to as the quantum connection. Below, we examine this connection 
from the viewpoint of coordinate wave functions.

To do this, we first choose a Poincar\'e 
gauge $\Bs$.  For any smooth curve $\gamma$ joining the points $x'$ and 
$x''$, the connection on $P\BM$ gives rise to an operator $\tau_\gamma(x'',
x') : \Pi^{-1}(x') \to \Pi^{-1}(x'')$ for parallel transport.  In turn, 
this defines the transformation
$g_\gamma^\Bs(x'',x') \in ISO_0(3,1)$ given by
\eqn\eggammadef{
\tau_\gamma(x'',x')\Bs(x') = \Bs(x'')g_\gamma^\Bs(x'',x'),}
where the term on the right side incorporates the (right) group action of
$ISO_0(3,1)$ on the fibre $\Pi^{-1}(x'') \subseteq P\BM$.  In view of the 
fact (cf.\ chapter 8 of [\rspivak]) that this action satisfies the 
compatibility condition $\tau_\gamma(x'',x')(\Bu'g) = \bigl(\tau_\gamma
(x'',x')\Bu'\bigr)g$ for any $\Bu' \in \Pi^{-1}(x')$ and any $g \in 
ISO_0(3,1)$, the transformations defined above satisfy an important 
composition law.  Namely, if $\gamma$ is smoothly extended to any third 
point $x''' \in \BM$, then we have the relation
\eqn\eggammacomp{
g_\gamma^\Bs(x''',x') = g_\gamma^\Bs(x''',x'')g_\gamma^\Bs(x'',x').}

In terms of the transformation defined in \eggammadef, the coordinate
wave function of any local quantum state in $\pi^{-1}(x') \subset \BE$
which is parallel transported from $x'$ to $x''$ along $\gamma$ is
then given by
\eqn\eggammaH{
{\it\Psi}_{\gamma,x''}^{\Bs(x'')}(q,p) = U\bigl(g_\gamma^\Bs(x'',x')\bigr) 
{\it\Psi}_{x'}^{\Bs(x')}(q,p), \quad\quad (q,p) \in {\cal P}_m^+,}
where $U$ is given by \ePoinrepH.
If we now denote the translation and Lorentz transformation parts of 
$g_\gamma^\Bs(x'',x')$ in \eggammadef\ as follows,
\eqn\eggammasplit{
g^\Bs_\gamma(x'',x') = \bigl(b^\Bs_\gamma(x'',x'),\,\Lambda^\Bs_\gamma
(x'',x')\bigr),}
then, according to \ePoinrepH, \eggammaH\ can be written in the form
\eqn\ebLambdaH{
{\it\Psi}_{\gamma,x''}^{\Bs(x'')}(q,p) =  {\it\Psi}_{x'}^{\Bs(x')}
\biggl(\Lambda_\gamma^\Bs(x'',x')^{-1}\bigl(q-b_\gamma^\Bs(x'',x')\bigr),
\,\,\Lambda_\gamma^\Bs(x'',x')^{-1}p\biggr).}

The connection constructed on $\BE$ enables us to generalize the 
special relativistic phase space propagators defined in \epropdef\ to
curved spacetimes.  Specifically, for any Poincar\'e gauge $\Bs$,
and any two points $x'$ and $x''$ joined by a smooth curve $\gamma$,
we define the propagator for parallel transport from $x'$ to $x''$ along 
$\gamma$ in the gauge $\Bs$ to be a function $K_\gamma^\Bs(x'',\zeta'';
x',\zeta')$ from ${\cal P}_m^+ \times {\cal P}_m^+ \to \BC$ given by 
[\rprugA,\rdrecA,\rprugB]
\eqn\egenpropdef{
K_\gamma^\Bs(x'',\zeta'';x',\zeta') = \inprod 
{\eta_{\zeta''}} {U\bigl(g_\gamma^\Bs(x'',x')\bigr)\eta_{\zeta'}}.}
In the above equation, we have introduced the composite
phase variable $\zeta = (q,p)$, so that the integration measure
$d\Sigma_m(q,p)$ on $\BF$ can be written simply as $d\Sigma_m(\zeta)$.
This notation will be employed throughout the rest of this section. 

In the case where the base manifold $\BM$ is flat, this propagator is
independent of the choice of $\gamma$.  It is then not difficult to show
[\rdrecA] that with an appropriate choice of section $\Bs$, the 
propagator for parallel transport assumes the same role as the quantum
propagator considered in section 2.  Namely, if we take for $\Bs$ 
the gauge in which $\Ba$ is identically ${\bf 0}$ and $\Be_i$ is
parallel to the positive $x^i$-axis, then the propagators defined
in \epropdef\ and \egenpropdef\ are related by
\eqn\eKrel{
K_\gamma^\Bs(x'',\zeta''\,;\,x',\zeta') = K(q''+x'',p''\,;\,q'+x',p'),
\quad \zeta'=(q',p'), \quad \zeta''=(q'',p'').}

With curved spacetimes the situation is more complicated.  In chapter
4 of [\rprugA], \prug\ proposes the idea of a quantum-geometric 
propagator, denoted by $\BK$.  This propagator is constructed in a similar 
manner as the conventional Feynman propagator, except that the space of 
paths which is integrated over is formed from arcs of geodesics in $\BM$ 
instead of the usual straight line segments. Physically, $\BK$ is to be 
interpreted as representing the evolution of a single-particle system
against a geometro-dynamic spacetime background (cf.\ section 4.7 of 
[\rprugA] for details), so that, in agreement with \eKrep, the 
quantum-geometric wave function
\eqn\eQdef{
{\it \Psi}_{x''}^{\Bs(x'')}(\zeta'') = \int_{\Sigma_m^+}\BK^\Bs(x'',\zeta'';
x',\zeta')\,{\it \Psi}_{x'}^{\Bs(x')}(\zeta')\,d\Sigma_m(\zeta'),}
represents the relative probability amplitude for the detection at $x''$ 
of the stochastic phase space value $\zeta''$ (relative to the gauge $\Bs$),
for a system prepared at $x'$ in the local state ${\bf \Psi}_{x'}$. 

To construct this propagator, we first foliate the given spacetime into
a family of Cauchy hypersurfaces $\sigma_t$ indexed by the global
time parameter $t$.  Choosing two hypersurfaces $\sigma_{t'}$ and 
$\sigma_{t''}$ with $t'' > t'$ and some positive integer $N$, we insert the 
$N-1$ intermediate hypersurfaces $\sigma_{t_n}$, where $t_n = {1 \over N}
\bigl[(N-n)t'+ nt'')\bigr]$ for $n = 1,2,\ldots,N-1$.  For any two points 
$x'$ and $x''$ lying on the surfaces $\sigma_{t'}$ and $\sigma_{t''}$ 
respectively, and for any two phase space points $\zeta'$ and $\zeta''$ 
(relative to the gauge $\Bs$), the quantum-geometric propagator 
$\BK^\Bs(x'',\zeta'';x',\zeta')$ is defined to be the limit 
[\rprugA,\rdrecA,\rprugB]
\eqn\eKKdefA{
\BK^\Bs(x'',\zeta'';x',\zeta') = \lim_{N \to \infty}\int\prod_{n=N}^1
K^\Bs_{\gamma(x_n,x_{n-1})}(x_n,\hat\zeta_n\,;\,x_{n-1},\hat\zeta_{n-1})\,
d\Sigma(x_n,p_n).}
Here, $\gamma(x_n,x_{n-1})$ is the geodesic joining $x_{n-1} \in 
\sigma_{t_{n-1}}$ to $x_n \in \sigma_{t_n}$, the measure 
on the hypersurfaces $\Sigma = \sigma \times \BV_m^+$ is given by
\eqn\eKKdefB{
d\Sigma(x_n,p_n)=2(p_n)^i\,d\sigma_i(x_n)\,d\Omega(p_n),}
and the intermediate phase points are chosen so that they correspond
with the point of contact between the tangent space and the base
manifold:
\eqn\eKKdefC{
\hat\zeta_n = \bigl(-a(x_n),p_n\bigr), \quad\quad a^i(x_n)\Be_i = \Ba(x_n),
\quad\quad n=1,2,\ldots,N-1.}
The points $x_0$ and $x_N$ correspond to the initial and final points
between which propagation takes place, i.e.\ we have 
$x_0=x'$, $x_N=x''$, $\hat\zeta_0=\zeta'$, and $\hat\zeta_N=\zeta''$.
Note also that the product in \eKKdefA\ does not include an integration
over $\sigma_{t''}$, so that only the intermediate hypersurfaces are
integrated over.

In the case of Minkowski space, and with the appropriate choice of gauge, it can be shown [\rprugA,\rdrecA,\rprugB] using \eKcomp\ and \eKrel\ that this 
quantity reduces to the propagator for parallel transport given in
\egenpropdef.
However, very little is known in the case where that spacetime is curved.
Clearly, a first step to understanding this problem is the evaluation
of the propagators for parallel transport contained in the product in
\eKKdefA.  It is this problem which we now consider.

Assuming as always that the resolution generator $\tilde\eta$ is taken
to be the standard one given in \emomresval, it follows from \epropdef\ that 
\egenpropdef\ can be written as
\eqn\egenpropcalc{
K_\gamma^\Bs(x'',\zeta'';x',\zeta') = 
\inprod {U(q'',\Lambda_{v''})\eta} {U(b+\Lambda q',\Lambda\Lambda_{v'})\eta},}
where $(b,\Lambda)=g_\gamma^\Bs(x'',x')$.  In order to evaluate this
expression, we use the fact [\rdrecA] that we can write the product
of Lorentz transformations in the inner product as $\Lambda\Lambda_{v'} =
\Lambda_{\Lambda v'}\Lambda_R$, where $\Lambda_R$ is a spatial rotation.  
Since $\eta$ is rotationally invariant (recall section 2), it 
follows from \epropdef\ and \epropval\ that
\eqn\egenpropval{
K_\gamma^\Bs(x'',\zeta'';x',\zeta') \,=\, {2\pi \over m^2 \tilde Z_{\ell,m}}
\,{K_1\bigl(m\,f_{\ell,m}(q,p)\bigr) \over f_{\ell,m}(q,p)},
\quad p = p'' + \Lambda p', \quad q = q''- \Lambda q'-b,} 
where $f_{\ell,m}$ is given by \efmldef.
The calculation of the propagator for parallel transport 
thus reduces to the problem of determining $b$ and $\Lambda$. 

As shown in [\rprugB], the 1-forms given in \eformsval\ can be
used to compute an explicit formula for $g^\Bs_\gamma(x'',x')$.
We shall now present that derivation in a form best suited to the 
computations in the subsequent two sections.

Suppose that $x'$ and $x''$ in $\BM$ are connected by a smooth curve
$\gamma$ with $\gamma(t')=x'$ and $\gamma(t'')=x''$.  Choose a positive
integer $n$ and divide $\gamma$ into $n$ pieces by defining the points
\eqn\edefpt{
x_k = \gamma(t_k), \quad\quad t_k = {(n-k)t' + kt'' \over n}, 
\quad\quad k=0,1,2,\ldots,n,} 
so that in particular $x_0 = x'$ and $x_n = x''$.  By induction on
\eggammacomp\ we have
\eqn\eggamman{
g^\Bs_\gamma(x'',x') = \prod_{k=n-1}^0g^\Bs_\gamma(x_{k+1},x_k).}
Applying the composition law for $ISO_0(3,1)$ to the right side and using 
\eggammasplit\ we see that for any $n=1,2,3,\ldots$, the total translation 
and total Lorentz transformation from $x'$ to $x''$ are respectively given by
\eqn\ebLambdagamman{
b = \sum_{k=0}^{n-1}\left[\prod_{l=n-1}^{k+1}
\Lambda_\gamma^\Bs(x_{l+1},x_l)\right]b_\gamma^\Bs(x_{k+1},x_k),\quad
\Lambda = \prod_{k=n-1}^0\Lambda_\gamma^\Bs
(x_{k+1},x_k).}

In order to compute these expressions, we use the exponential map from
$iso(3,1)$ to $ISO_0(3,1)$.  Specifically, let $\Delta t = (t'' - t')/n$
be the increment in the parameter $t$ corresponding to \edefpt.  Then to
the first order in $\Delta t$ we have
\eqn\eapproxggammainc{
b_\gamma^\Bs(x_{k+1},x_k) \approx \Delta t\,P^\Bs(\BX_k), \quad\quad
\Lambda_\gamma^s(x_{k+1},x_k) \approx 
\exp\bigl(\Delta t\,M^\Bs(\BX_k)\bigr),}
where $\BX_k = \gamma'(t_k)$ is the tangent vector to $\gamma$ at $t = t_k$,
and the infinitesimal translation and Lorentz transformation are defined
in terms of the standard basis elements $P_i$ and $M^{ij}$ of $iso(3,1)$, 
and the 1-forms given in \eformsval, by
\eqn\ePMdef{
P^\Bs(\BX) = -\theta_\Bs^i(\BX)P_i,\quad\quad 
M^\Bs(\BX) = -{1 \over 2} \omega_{ij}^\Bs (\BX)M^{ij},\quad\quad
\BX \in T\BM.}
Substituting \eapproxggammainc\ into \ebLambdagamman\ gives the
approximations
\eqn\ebLambdaapprox{
b \approx \Delta t\sum_{k=0}^{n-1}\left[\prod_{l=n-1}^{k+1}
\exp\bigl(\Delta t\,M^\Bs(\BX_l)\bigr)\right]
P^\Bs(\BX_k),\quad\!
\Lambda \approx \prod_{k=n-1}^0\exp\bigl(\Delta t\,M^\Bs(\BX_k)\bigr).}
In the limit as $n \to \infty$ the approximations will become exact, and
we have as our final answer
\eqn\ebgamma{
b = \int_{t'}^{t''}\left[T\,\exp\left(\int_{t''}^t
M^\Bs\bigl(\gamma'(s)\bigr)\,ds\right)
P^\Bs\bigl(\gamma'(t)\bigr)\right]\,dt,}
\eqn\eLambdagamma{
\Lambda = T\,\exp\left(\int_{t''}^{t'}M^\Bs
\bigl(\gamma'(t)\bigr)\,dt\right).}

\newsec {Quantum parallel transport in Robertson-Walker spacetimes}

As shown in section 3, the propagators for parallel transport are
completely determined by the quantities $b$ and $\Lambda$ given in
\ebgamma\ and \eLambdagamma.  An example of the application of these
formulae is provided by the classical Robertson-Walker model, whose
metric can be expressed [\rschutz] in the form
\eqn\erwmet{
ds^2 = -dt^2 + a^2(t)
\left[ {dr^2 \over 1-kr^2} + r^2 d\Omega^2 \right], \quad
d\Omega^2 = d\theta^2 + \sin^2 \theta\,d\phi^2,}
where $a(t)$ is a smooth positive function of $t$.
The parameter $k$ can be either $-1$, $0$, or $1$, corresponding respectively
to spatially hyperbolic, spatially flat, and spatially elliptic 
universes.  In this section, we compute the geodesics in the
Robertson-Walker model (employing a method suggested by \prug\ [\rprugC]), 
and use them to arrive at integral representations for $b$ and $\Lambda$.

To begin, it is convenient to define a new variable $R$ by
\eqn\ertoR{
R = \cases{\arcsinh r, &if $k=-1$;\cr
r, &if $k=0$;\cr
\arcsin r, &if $k=1$,\cr}}
which reduces the metric to the form
\eqn\erwmetnew{
ds^2 = -dt^2 + a^2(t)\bigl[dR^2 + r^2 d\Omega^2\bigr].}

For the calculation of the geodesics, we concentrate on curves
whose spatial projections follow paths passing through the origin.
Clearly, any geodesic can be obtained from one of this form by a
suitable shift of the origin.
Along such lines $\theta$ and $\phi$ are constant, and so the distance 
along a curve $R(t),\,t' \le t \le t''$ can be expressed as
\eqn\edist{
l = \int_{t'}^{t''}
{\cal L}(R,{\dot R},t)\,dt, \quad\quad
{\cal L}(R,{\dot R},t)=\sqrt{\pm\bigl(1-a^2(t)\dot{R}^2\bigr)},}
where the plus sign is taken for timelike geodesics and the minus sign
for spacelike geodesics.

To minimize or maximize $l$ as a functional of 
the path we use the Euler-Lagrange equation, which in the case of 
timelike geodesics takes the form [\rdods]
\eqn\eeuler{
0={\partial {\cal L} \over \partial R} - 
{d \over dt} {\partial {\cal L} \over \partial {\dot R}}= {d \over dt} \left(
{a^2(t) {\dot R} \over \sqrt{1-a^2(t){\dot R}^2}} \right).}
Integrating with respect to $t$ gives
\eqn\ertime{
R(t) = C\int {dt \over a(t) \sqrt{C^2+a^2(t)}},}
where $C \ge 0$ is a constant of integration.
A similar computation for the spacelike case yields
\eqn\erspace{
R(t) = C\int {dt \over a(t) \sqrt{C^2-a^2(t)}}.}

We now turn to the problem of calculating $b$ and $\Lambda$.
For simplicity, we adopt a Poincar\'e gauge in which the
displacement vector $\Ba$ is identically $\B0$.  From \erwmetnew\ 
it follows that a natural choice for the basis vectors is given by
\eqn\ebasis{
\Be_0=\partial_t,\quad\; \Be_1={1 \over a(t)}\partial_R, 
\quad\; \Be_2={1 \over r\,a(t)}\partial_\theta,\quad\;
\Be_3={1\over r\,a(t)\sin\theta}\partial_\phi.}
Their Lie brackets are then computed to be 
\eqn\ecomm{
{\eqalign{[\Be_0,\Be_1] = -{a'(t) \over a(t)}\Be_1, \quad
&[\Be_0,\Be_2] = -{a'(t) \over a(t)}\Be_2, \quad
[\Be_0,\Be_3] = -{a'(t) \over a(t)}\Be_3, \cr
[\Be_1,\Be_2] = -{1 \over r\,a(t)}{dr \over dR} \Be_2, \quad\,
&[\Be_1,\Be_3] = -{1 \over r\,a(t)}{dr \over dR} \Be_3, \quad\,
[\Be_2,\Be_3] = -{\cot \theta \over r\,a(t)} \Be_3.}}}
{}From \eformsval, the nonzero coefficients corresponding to Lorentz boosts 
are found to be
\eqn\eRWboost{
\omega_{01}^\Bs(\Be_1) = \omega_{02}^\Bs(\Be_2) = \omega_{03}^\Bs(\Be_3) = 
{a'(t) \over a(t)},}
while for spatial rotations they are
\eqn\eRWrot{
\omega_{12}^\Bs(\Be_2) = \omega_{13}^\Bs(\Be_3) = 
{1 \over r\,a(t)}{dr \over dR}, \quad\quad
\omega_{23}^\Bs(\Be_3) = {\cot \theta \over r\,a(t)}.}
Since the displacement vector $\Ba$ is assumed to be identically $\B0$, it
follows from \eformsval\ that the 1-forms for translation will simply
be given by $\theta_\Bs^i(\BX) = X^i$.

For the computations of $b$ and $\Lambda$ we will take $\gamma$ to be 
one of the special geodesics considered in \edist.  Because of the similarity
of \ertime\ and \erspace, it is convenient in the following calculations
to consider the cases of timelike and spacelike geodesics together.
Following the example of \edist, we will adopt the convention that
whenever a $\pm$ sign occurs the plus refers to the timelike case and the
minus to the spacelike case.

Let the initial and final points of $\gamma$ be $x'$ and $x''$, 
corresponding to the global time values $t'$ and $t''$ respectively.
{}From \ertime\ and \erspace\ the tangent vector is seen to be
\eqn\eRWtangent{
\gamma'(t) = \Be_0 + {C \over c(t)}\Be_1, 
\quad\quad c(t) = \sqrt{C^2 \pm a^2(t)}.}  
It then follows from \ePMdef\ and \eRWboost-\eRWtangent\  
that the infinitesimal translation and Lorentz transformation elements of
$iso(3,1)$ along $\gamma$ are respectively
\eqn\ePMgeo{
P^\Bs\bigl(\gamma'(t)\bigr) = -P_0 - {C \over c(t)}P_1, \quad\quad
M^\Bs\bigl(\gamma'(t)\bigr) = -{C a'(t) \over a(t)c(t)}M^{01}.}
Clearly the quantities $M^\Bs\bigl(\gamma'(t)\bigr)$ commute for all values 
of $t$, and so \ebgamma\ and \eLambdagamma\ become
\eqn\ebgammaRW{
b = -\int_{t'}^{t''}\left[\exp\left(-\int_t^{t''}
{Ca'(s) \over a(s)c(s)}M^{01}\,ds\right)
\left(1,{C \over c(t)},0,0\right)\right]\,dt,}
\eqn\eLambdagammaRW{
\Lambda = \exp\left(-\int_{t'}^{t''}
{C a'(t) \over a(t)c(t)}M^{01}\,dt\right).}

The case $a(t)=1$ corresponds to Minkowski space.  Here,
\ebgammaRW\ and \eLambdagammaRW\ reduce to the equations
\eqn\eRWspecialA{
b = (t'-t'')\left[P_0 + {C \over \sqrt{C^2 \pm 1}}P_1\right],
\quad\quad \Lambda = 1.}
In this case it follows from \ebLambdaH\ that the coordinate wave functions 
at the endpoints are related by
\eqn\eRWspecialB{
{\it\Psi}_{\gamma,x''}^{\Bs(x'')}(q,p) = 
{\it\Psi}_{x'}^{\Bs(x')}\left(q + (t''-t')\left[P_0 + {C \over 
\sqrt{C^2 \pm 1}} P_1\right],\,p\right),}
confirming the fact that two observers in flat space at rest relative to one 
another and with coincident axes will measure the same wave function 
for the system, modulo a spacetime translation.

\newsec {Explicit propagators for a spatially-flat Robertson-Walker model}

The integral representations for $b$ and $\Lambda$
which were derived in section 4 generally cannot be evaluated in
closed form.  Indeed, even in the spatially-flat case corresponding to $R=r$,
the determining equations for $C$ 
which are obtained from \ertime\ or \erspace\ cannot be solved explicitly, 
except in special cases.  In this section we consider one such case, 
the rapidly-expanding universe corresponding to $a(t)=t$.

We begin by choosing coordinates $(t,x^1,x^2,x^3)$ for $\BM$, in terms of 
which the metric assumes the form $ds^2 = -dt^2+t^2\bigl((dx^1)^2+
(dx^2)^2+(dx^3)^2\bigr)$, and we 
take as our Poincar\'e gauge a section $\Bs$ of $P\BM$ in which the
displacement vector $\Ba$ is identically $\B0$ and the orthonormal frame is
given by
\eqn\ebasist{
\Be_0 = \pd {} t, \quad\quad \Be_1 = {1 \over t}\pd {} {x^1}, \quad\quad
\Be_2 = {1\over t} \pd {} {x^2}, \quad\quad \Be_3 = {1\over t}\pd {} {x^3}.}
The initial and final points of the geodesic
are taken to be $x'=(t',\Bx')$ and $x''=(t'',\Bx'')$, with $t'' > t'$.
We decompose the spatial separation of the two points as $\Bx''-\Bx'=
\rho\Bn$, where $\rho \ge 0$ and $\Bn = (n_1,n_2,n_3)$ is a unit vector. 
For future reference we define the quantities $\rhonull$ and $\rhosl$ by
\eqn\edefs{
\rhonull = \ln{t'' \over t'}, \quad\quad \rhosl = \ln
\left({t'' + \sqrt{(t'')^2-(t')^2} \over t'}\right).}

Consider first the case where the geodesic joining $x'$ to $x''$ is
timelike.  We begin by calculating the constant of integration in \ertime,
which for definitiveness will be denoted as $\Ctl$.  For this purpose it is 
convenient to define the new variable $u = \sqrt{t^2+\Ctl^2}$, with
$u'$ and $u''$ defined accordingly.  Using \ertime\ with $a(t)=t$ we get
\eqn\ecalcCtl{
\rho = \int_{t'}^{t''}{\Ctl \over t \sqrt{\Ctl^2+t^2}}\,dt = 
{1 \over 2}\ln\left({(u'+\Ctl)(u''-\Ctl) \over (u'-\Ctl)(u''+\Ctl)}\right).}
Exponentiating and squaring this equation gives the pair of relations
\eqn\ecalcCnexttl{
(u'+\Ctl)(u''-\Ctl)=e^\rho t't'', \quad\quad (u'-\Ctl)(u''+\Ctl)=
e^{-\rho} t't'',}
which when added together lead to the result
\eqn\eCvaltl{
\Ctl = {e^{-\rho/2}(e^{2\rho}-1)t't'' \over 2\sqrt{(t''-e^\rho t')
(e^\rho t''-t')}}, \quad\quad 0 \le \rho < \rhonull.}

We next proceed to the calculation of $\Lambda$.  To simplify this 
computation, it is convenient to first consider the case in which
$x^2$ and $x^3$ are constant along $\gamma$, so that $\Bn=(1,0,0)$.  In 
this case the Lorentz transformation is found from \eLambdagammaRW\ to be
\eqn\eLambdatl{
\Lambda = \exp\left(-\int_{t'}^{t''}
{\Ctl \over t \sqrt{\Ctl^2+t^2}}M^{01}\,dt\right), 
\quad\quad \Bn=(1,0,0).}
A comparison of this formula with \ecalcCtl\ shows that $\Lambda$ is the 
standard Lorentz boost in the $x^1$ direction, with $\cosh\rho$ in its 
$(0,0)$ and $(1,1)$
entries and $-\!\sinh\rho$ in its $(0,1)$ and $(1,0)$ entries.  This result 
can then be extended to arbitrary values of $\Bn$ by noting that since the
spacetime in question is spatially flat, it is only necessary to
conjugate $\Lambda$ with an appropriate spatial rotation matrix which maps 
$(1,0,0)$ into $\Bn$.  In this manner the Lorentz transformation for 
a general geodesic $\gamma$ is found to be
\eqn\eLambdavalt{
\Lambda = \left[\matrix {\cosh\rho& -n_1\sinh\rho & -n_2\sinh\rho & 
-n_3\sinh\rho \cr
-n_1\sinh\rho & n_1^2(\cosh\rho-1)+1 & n_1 n_2(\cosh\rho-1) & 
n_1 n_3(\cosh\rho-1) \cr
-n_2\sinh\rho & n_2 n_1(\cosh\rho-1) & n_2^2(\cosh\rho-1)+1 & 
n_2 n_3(\cosh\rho-1) \cr
-n_3\sinh\rho & n_3 n_1(\cosh\rho-1) & n_3 n_2(\cosh\rho-1) & 
n_3^2(\cosh\rho-1)+1 \cr}
\right].}

To calculate the translation $b$ provided by \ebgammaRW, note that due to the 
spatial flatness of $\BM$ the spatial part of $b$ will be parallel to the 
direction of propagation.  Thus, if we denote the temporal and spatial parts
of $b$ by $b_T$ and $b_S$, then $b$ can be written in the form
$b = (b_T,\,b_S \Bn)$.  In order to compute
these quantities, we again consider first the case of
propagation along the $x^1$-axis, for which $\Bn = (1,0,0)$, and obtain from 
\ebgammaRW\ the formula
\eqn\ebtl{
b = -\int_{t'}^{t''}\left[\exp\left(-\int_t^{t''}
{\Ctl \over s \sqrt{\Ctl^2+s^2}}M^{01}\,ds\right)
\left(1,{\Ctl \over \sqrt{\Ctl^2+t^2}},0,0\right)\right]\,dt.}
Using \ecalcCtl\ to evaluate the matrix integral inside the exponential, 
we get by equating the first two components of the left and right sides
of \ebtl\ the equations 
\eqn\ebcalct{
{\eqalign{
&b_T = -\int_{t'}^{t''}\left[\cosh(\ln v) - {\Ctl \sinh(\ln v) 
\over u}\right] dt, \cr
b_S = -\int_{t'}^{t''}&\left[
{\Ctl \cosh(\ln v) \over u} - \sinh(\ln v)\right] dt, \quad  
v = {\sqrt{u+\Ctl}\sqrt{u''-\Ctl} \over \sqrt{u-\Ctl}\sqrt{u''+\Ctl}}.}}}
These integrals are most easily computed by expressing them in a form
in which the limits are $u'$ and $u''$.  In the case of $b_T$, this change
of variables produces an elementary integral,
\eqn\ebTvaltl{
b_T \,\,=\,\, {1 \over 2}\int_{u'}^{u''}{\Ctl(v^2-1)-u(v^2+1) \over 
v\sqrt{u^2-\Ctl^2}}\,du \,\,=\,\, -{u'' \over t''}\int_{u'}^{u''}\!\! du 
\,\,=\,\, {u''(u'-u'') \over t''},}
while for $b_S$ a similar calculation gives
\eqn\ebSvaltl{
b_S \,\,=\,\, {1 \over 2}\int_{u'}^{u''}{u(v^2-1)-\Ctl(v^2+1) \over
v\sqrt{u^2-\Ctl^2}}\,du \,\,=\,\, -{\Ctl \over t''}\int_{u'}^{u''}\!\! du
\,\,=\,\, {\Ctl(u'-u'') \over t''}.}
Substituting $u=\sqrt{t^2+\Ctl^2}$ into \ebTvaltl\ and \ebSvaltl\ and
using \eCvaltl\ to eliminate $\Ctl$ gives a pair of expressions
involving $e^\rho$.  These can be simplified by converting the exponentials
into hyperbolic functions, and the final result is then found to be
\eqn\ebvals{
b_T = t'\cosh\rho-t'', \quad\quad\quad b_S = -t'\sinh\rho.}
Using the spatial flatness of $\BM$, it is not difficult to see that these
formulae in fact hold for all values of the direction vector $\Bn$.

The computations performed above can be repeated for the spacelike case, with 
similar results.  However, one important difference arises.  To see this, 
consider first the determination of the constant of integration in \erspace, 
denoted this time by $\Csl$.  Substitution of $a(t)=t$ into \erspace\ gives
\eqn\ecalcCsl{
\rho = \int_{t'}^{t''}{\Csl \over t \sqrt{\Csl^2-t^2}}\,dt.}  
A comparison of this equation with \ecalcCtl\ immediately gives the
solution $\Csl = i\Ctl$, with $\Ctl$ given by \eCvaltl.  Note that $\Csl$ 
is actually real since in the case of spacelike geodesics we have 
$\rho > \rhonull$ and hence $t''-e^\rho t' < 0$.
However, it follows also from \ecalcCsl\ that the range of $\Csl$ 
will be restricted to the interval $[t'',\infty)$.
The value $\Csl=t''$ corresponds to the greatest possible spatial separation
of the points $x'$ and $x''$, which is achieved by a geodesic with points
given by $\{(t,\Bx'+\rmax(t)\Bn)\,\vert\,t'\le t \le t''\}$, where
\eqn\egeomax{
\rmax(t) = \int_{t'}^t {t'' \over s\sqrt{(t'')^2-s^2}}ds =
{1 \over 2}\ln \left[{\bigl(t''+\sqrt{(t'')^2-(t')^2}\bigr)
\bigl(t''-\sqrt{(t'')^2-t^2}\bigr) \over
\bigl(t''-\sqrt{(t'')^2-(t')^2}\bigr)
\bigl(t''+\sqrt{(t'')^2-t^2}\bigr)}\right].}
Substitution of $t=t''$ into this equation and comparison with the
second equation in \edefs\ reveals that such a geodesic will achieve 
a spatial separation of magnitude $\rhosl$ between $x'$ and $x''$.
Consequently, geodesics connecting two points with a spatial
separation greater that $\rhosl$ are formed by smoothly extending the
geodesic given by \egeomax\ into a straight line with spatial projection
parallel to $\Bn$, lying in the Cauchy hypersurface $\Sigma_{t''}=
\{(t'',\Bx)\,\vert\,\Bx \in \BR^3\}$.  This extension is easily seen to
satisfy the geodesic equation at every point.

Let us first consider the geodesics for which $\rhonull < \rho
\le \rhosl$.  In this case the translation and Lorentz transformation
for geodesics whose spatial projection is parallel to the $x^1$ axis 
are found from \ebgammaRW\ and \eLambdagammaRW\ to be
\eqn\ebsl{
b = -\int_{t'}^{t''}\left[\exp\left(-\int_t^{t''}
{\Csl \over s \sqrt{\Csl^2-s^2}}M^{01}\,ds\right)
\left(1,{\Csl \over \sqrt{\Csl^2-t^2}},0,0\right)\right]\,dt,}
\eqn\eLambdasl{
\Lambda = \exp\left(-\int_{t'}^{t''}
{\Csl \over t \sqrt{\Csl^2-t^2}}M^{01}\,dt\right).}
Substitution of $i\Ctl$ for $\Csl$ in the above expressions and comparison 
with \eLambdatl\ and \ebtl\ shows that the formulae for $b$ and $\Lambda$ in 
the spacelike case formally reduce to those for the timelike case.

For geodesics $\gamma$ corresponding to $\rho \ge \rhosl$ we define an 
intermediate point $\bar x = (\bar t, \bar \Bx)$ on $\gamma$ by 
$\bar t = t''$ and $\bar \Bx = \Bx' + \rhosl \Bn$, so that as $x$ varies from
$x'$ to $\bar x$, $\gamma$ follows the geodesic given by \egeomax, while as
$x$ varies from $\bar x$ to $x''$, $\gamma$ follows a straight line along 
the hypersurface $\Sigma_{t''}$.  It then follows from the composition law 
for $ISO_0(3,1)$ that the translation and Lorentz transformation from 
$x'$ to $x''$ are given by 
\eqn\ecomplawA{
\Lambda = \Lambda_\gamma^\Bs(x'',\bar x)\Lambda_\gamma^\Bs(\bar x,x'), 
\quad\quad b = b_\gamma^\Bs(x'',\bar x) + \Lambda_\gamma^\Bs(x'',\bar x)
b_\gamma^\Bs(\bar x,x').}
The quantity $\Lambda_\gamma^\Bs(x'',\bar x)$ can be obtained from 
\eLambdagamma\ by considering $t$ to represent spatial instead of temporal
distance, and integrating from $\rhosl$ to $\rho$ along the tangent vector 
$t''\Be_1$.  An easy computation then shows that $\Lambda_\gamma^\Bs (x'',
\bar x)$ is a matrix of the form \eLambdavalt\ with $\cosh\rho$ and $\sinh
\rho$ replaced by $\cosh(\rho-\rhosl)$ and $\sinh(\rho-\rhosl)$ respectively.
Consequently, it follows from the first equation in \ecomplawA\ that the 
total Lorentz transformation from $x'$ to $x''$ is given as before 
by \eLambdavalt.

To calculate the total translation, we decompose $b_\gamma^\Bs(\bar x,x')$ 
and $b_\gamma^\Bs (x'',\bar x)$ into temporal and spatial parts in the 
earlier-mentioned manner.  In terms of this decomposition, the second 
equation in \ecomplawA\ can be expressed in the matrix form
\eqn\ecomplawB{
\left[\matrix{b_T \cr b_S}\right] = 
\left[\matrix{b_T(x'',\bar x) \cr b_S(x'',\bar x)}\right]+
\left[\matrix{\cosh(\rho-\rhosl) & -\sinh(\rho-\rhosl) \cr
-\sinh(\rho-\rhosl) & \cosh(\rho-\rhosl)}\right]
\left[\matrix{b_T(\bar x,x') \cr b_S(\bar x,x')}\right].}
To calculate $b_T(\bar x,x')$ and $b_S(\bar x,x')$, first note that from
\edefs\ we have
\eqn\etrigrho{
\cosh\rhosl = {t'' \over t'}, \quad\quad\quad
\sinh\rhosl = {\sqrt{(t'')^2-(t')^2} \over t'}.}
Substituting $\rho = \rhosl$ into \ebvals\ and applying the first formula 
above then gives 
\eqn\ebvalfirstpart{
b_T(\bar x,x') = 0, \quad\quad 
b_S(\bar x,x') = -t'\sinh\rhosl.}
Using \ebgamma, the quantities $b_T(x'',\bar x)$ and $b_S(x'',\bar x)$ can be 
evaluated by a calculation similar to the derivation of $\Lambda_\gamma^\Bs
(x'',\bar x)$ as
\eqn\ebvalsecondpart{
b_T(x'',\bar x) = t''[\cosh(\rho-\rhosl)-1], \quad
b_S(x'',\bar x) = -t''\sinh(\rho-\rhosl),}
and so we get by substituting \ebvalfirstpart\ and \ebvalsecondpart\ into 
\ecomplawB\ that for $\rho \ge \rhonull$ the spacetime translation is
given by
\eqn\ebdistant{
{\eqalign{
b_T &= t''[\cosh(\rho-\rhosl)-1]+t'\sinh(\rho-\rhosl)
\sinh\rhosl,\cr
b_S &= -t''\sinh(\rho-\rhosl)-t'\cosh(\rho-\rhosl)\sinh\rhosl.}}}
Expanding out the hyperbolic functions in these expressions and then using
\etrigrho, we find that once again they reduce formally to \ebvals.  As
before, this result is easily seen to hold for all choices of the
direction vector $\Bn$.

So far, we have not made any mention of the case where the geodesic 
joining $x'$ to $x''$ is null.  Although this case could be handled 
directly using \ebgamma\ and \eLambdagamma, it is more straightforward to note 
that since $b$ and $\Lambda$ are continuous functionals of the path
$\gamma$, these quantities will be given by the formulae already derived.

In conclusion, the expressions in \eLambdavalt\ and \ebvals\ provide, through 
\efmldef\ and \egenpropval, an explicit form of the propagator for 
parallel transport between any two points $x'$ and $x''$ in $\BM$ for 
which $t'' > t'$.
In the concluding paragraph of [\rprugB], \prug\ speculates that the
quantum-geometric propagators associated with rapidly expanding or
contracting universes may lead to violations of strict Einstein
causality.  The calculations presented above can serve as a basis
for the investigation of such phenomena.

\ack{
The author would like to thank E. \prug\ for discussions and suggestions
during the preparation of this paper.  He would also like to thank the
referees for suggestions for improvement in the exposition of background
material.  The Maple computer algebra system, a product of Waterloo Maple 
Software, was used for several of the computations in section 5.}

\immediate\closeout\rfile%
\bigskip\noindent{\bf References}\par\nobreak
\frenchspacing\input refs.tmp
\end